\documentclass[12pt]{article}
\usepackage{amsmath,amsfonts,amssymb,latexsym}

\def\be{\begin{equation}}
\def\ee{\end{equation}}
\def\bq{\begin{eqnarray}}
\def\eq{\end{eqnarray}}
\def\beq{\begin{eqnarray*}}
\def\eeq{\end{eqnarray*}}

\begin{document}

\title{A General Sudden Cosmological Singularity}
\author{John D. Barrow$^{1}$, S. Cotsakis$^{2}$ and A. Tsokaros$^{3}$ \\
$^{1}$DAMTP, Centre for Mathematical Sciences, \\
Cambridge University, Cambridge CB3 0WA, UK \\
$^{2,3}$Research Group of Geometry, Dynamical Systems and Cosmology,\\
Dept. Information \& Communication Systems Engineering, \\
University of the Aegean, Karlovassi 83200, Samos, Greece.}
\maketitle

\begin{abstract}
We construct an asymptotic series for a general solution of the Einstein
equations near a sudden singularity. The solution is quasi isotropic and
contains nine independent arbitrary functions of the space coordinates as
required by the structure of the initial value problem.

PACS number: 98.80.-k
\end{abstract}

\section{Introduction}

The standard homogeneous and isotropic Friedmann-Lema\^{\i}tre cosmological
models of general relativity are determined by the solution of two
independent Einstein equations for three unknown time-dependent functions,
the metric scale factor $a(t)$, the fluid density, $\rho (t)$, or the
isotropic fluid pressure, $p(t)$. If an equation of state $p=f(\rho )$ is
chosen, the system closes and the two remaining unknown functions are
determined. Physically reasonable equations of state, like those for perfect
fluids, produce well-behaved expansion factors with behaviours that offer
simple Newtonian interpretations. However, the challenge of providing a
compelling explanation for the observed acceleration of the universe in
terms of a 'dark energy' fluid has led to an exploration of other less
familiar equations of state \cite{jdb3}, motivated by the form of bulk
viscous stresses \cite{visc,visc2,visc3}, where $dp/d\rho $ is not
everywhere continuous, or scenarios in which there is no equation of state
at all. These less constrained material stresses introduce new possibilities
for the evolution of Friedmann universes and can produce unexpected types of
finite-time singularity which are far weaker than the curvature
singularities that often accompany geodesic incompleteness \cite%
{tipl,kr,laz,laz2,bal,HE}. We found \cite{jdb04a,jdb04aa,jdb04b} a whole new
class of pressure-driven singularities that keep the scale factor, $a$, the
cosmological expansion rate, $\dot{a}/a$, and the density, $\rho $, finite
while permitting the pressure, $p$, and the deceleration $-\ddot{a}/a$, to
blow up at a finite time despite the energy conditions $\rho >0$ and $\rho
+3p>0$ holding, and regardless of the three-curvature. There is no geodesic
incompleteness associated with these singularities.

The stability of general relativistic solutions with sudden singularities in
the presence of small scalar, vector, and tensor perturbations has been
demonstrated in a gauge covariant formalism \cite{lip}, and they have also
been shown to be stable against quantum particle production processes as the
singularities is approached \cite{fab}. They have been investigated in a
number of different cosmological setting and their behaviour has been
classified in the light of other types of finite-time singularity that can
arise in isotropic and anisotropic cosmologies \cite%
{bt,rev,rev1,rev2,sh,dab,noj,odin,cat,dab2,dab3}. The formal definitions of
these singularities have been discussed by Lake \cite{lake} in the case
where infinities occur in second derivatives of $a(t)$, but similar examples
exist where the sudden singularity occurs in (arbitrarily) higher
derivatives of $a(t)$ and so no energy conditions are threatened \ \cite%
{jdb04b,bt}. Recently, we have also provided another formal construction
procedure for these solutions in isotropic universes using fractional power
series \cite{scot,ant}.

Sudden singularities have appeared in a wide range of cosmological
investigations and our aim in this paper is to try to obtain information
about the general cosmological behaviour in the neighbourhood of such a
singularity by an asymptotic analysis of the Einstein equations without
using simplifying assumptions of homogeneity or isotropy. We shall use a
function-counting method, familiar from ref \cite{LL}. Past uses of this
method have failed to identify general behaviour near isotropic
singularities \cite{LL,KL,KL2} because of the ubiquity of chaotic behaviour
near curvature singularities. However, although the full complement of
independently arbitrary functions did not appear, the series expansions
obtained provided useful asymptotic forms for particular cosmological models
containing fluids and combinations of fluids \cite{starob,starob2}. The only
use of this technique which identified a general behaviour in the Einstein
cosmological equations is the study of the late-time behaviour of expanding
universes with $\rho >0$ and $\rho +3p>0$ in the presence of a positive
cosmological constant. A series expansion around the de Sitter solution \cite%
{no hair} produces a general complement of arbitrary functions in accord
with the expectations of the cosmic no hair theorem \cite{no
hair,wald,jdb,jen,rend}.

In this paper we are going to seek a series expansion around the particular
FRW solution with a future sudden singularity introduced in ref \cite%
{jdb04a,jdb04aa}. In the limit that the sudden singularity is approached at
finite future time we will determine the number of independently arbitrary
functions of the three space variables that are necessary and sufficient to
prescribe the solution locally on a spacelike hypersurface of constant
comoving proper time.

\section{Friedmann-Lema\^{\i}tre sudden singularities}

Consider the most general form of sudden finite-time singularity of the sort
discussed in \cite{jdb04a,jdb04aa} that can arise in an Friedmann expanding
universe. In \cite{jdb04a} we showed that singularities of this sort can
arise from a divergence of the pressure, $p$, at finite time despite the
scale factor, $a(t)$, the material density, $\rho ,$ and the Hubble
expansion rate, $H=\dot{a}/a$ remaining finite. The pressure singularity is
accompanied by a divergence in the acceleration of the universe, $\ddot{a}$,
at finite time. Remarkably, these singularities occur without violating the
strong energy conditions $\rho >0$ and $\rho +3p>0.$ They can even prevent a
closed Friedmann universe that obeys these energy conditions from attaining
an expansion maximum \cite{jdb04aa}. In order to prevent the occurrence of
these sudden singularities it is necessary to bound the pressure. A
sufficient condition is to require $dp/d\rho $ to be continuous or $p/\rho $
to be finite.

We will construct a quasi-isotropic ansatz for the general behaviour by
first considering the Friedmann universe with curvature parameter $k$, zero
cosmological constant, and units $8\pi G=c=1$. The Einstein equations reduce
to

\begin{eqnarray}
3H^{2} &=&\rho -\frac{3k}{a^{2}},  \label{fried} \\
\dot{\rho}+3H(\rho +p) &=&0,  \label{con} \\
\frac{\ddot{a}}{a} &=&-\left( \frac{\rho +3p}{6}\right) .  \label{fr}
\end{eqnarray}%
In \cite{jdb04a} we constructed an explicit example by seeking, over the
time interval $0<t<t_{s},$ a solution for the scale factor $a(t)$ of the form

\begin{equation}
a(t)=1+Bt^{q}+C(t_{s}-t)^{n},  \label{ex}
\end{equation}%
where $B>0,q>0,C$ and $n>0$ are free constants to be determined. If we fix
the zero of time by requiring $a(0)=0,$ so $Ct_{s}^{n}=-1,$ we have

\begin{equation}
a(t)=(\frac{t}{t_{s}})^{q}\left( a_{s}-1\right) +1-(1-\frac{t}{t_{s}})^{n},
\label{sol2}
\end{equation}%
where $a_{s}\equiv a(t_{s})$. Hence, as $t\rightarrow t_{s}$ from below, we
have

\begin{equation}
\ddot{a}\rightarrow q(q-1)Bt^{q-2}-\frac{\ n(n-1)}{t_{s}^{2}(1-\frac{t}{t_{s}%
})^{2-n}}\rightarrow -\infty ,  \label{Lim}
\end{equation}%
whenever $1<n<2$ and $0<q\leq 1$; the solution exists on the interval $%
0<t<t_{s}$. Hence, as $t\rightarrow t_{s\text{ }}$we have $a\rightarrow
a_{s} $, $H\rightarrow H_{s}$ and $\rho \rightarrow \rho _{s}>0$ where $%
a_{s},H_{s},$ and $\rho _{s}$ are all finite but $p_{s}\rightarrow \infty $.
The infinite deceleration in (\ref{Lim}) is why this sometimes also termed a
'big brake' singularity. By contrast, as $t\rightarrow 0$ there is an
initial all-encompassing strong-curvature singularity, with $H\rightarrow
\infty ,\rho \rightarrow \infty $ and $p\rightarrow \infty $. From (\ref{Lim}%
) and (\ref{fr}), we see that $\rho $ and $\rho +3p$ remain positive
throughout the evolution but, because $\rho $ is finite asymptotically, the
dominant-energy condition, $\left\vert p\right\vert \leq \rho ,$ must always
be violated as $t\rightarrow t_{s}$, \cite{jdb04a,lake,bt}.

Using the isotropic and homogeneous solution (\ref{sol2}) as a guide we
consider a series expansion of the metric in the neighbourhood of the sudden
singularity in a general cosmological solution to the Einstein equations.

\section{General series expansions near the singularity}

\subsection{Fractional power series}

We use the Landau-Lifshitz \cite{LL} notation: Latin indices for spacetime
components, Greek for spatial components. The general form of the metric in
synchronous coordinates is
\begin{equation}
ds^{2}=dt^{2}-\gamma _{\alpha \beta }dx^{\alpha }dx^{\beta }.  \label{met}
\end{equation}%
On approach to the sudden singularity as $t\rightarrow t_{s}$, the solution (%
\ref{sol2}) has the linear asymptotic form

\begin{equation}
a\rightarrow a_{s}+q(1-a_{s})(1-\frac{t}{t_{s}}).  \label{lin}
\end{equation}

Guided by this FRW example, (\ref{sol2}), and its asymptotic form, (\ref{lin}%
), we carry out a change of time coordinate to place the future sudden
singularity at $t=0$ and then consider a general series expansion about it,
with

\begin{equation*}
\gamma _{\alpha \beta }=a_{_{\alpha \beta }}+b_{_{\alpha \beta
}}t+c_{_{\alpha \beta }}t^{n}+...,
\end{equation*}%
where $1<n<2$, and $a_{\alpha \beta },b_{\alpha \beta },c_{\alpha \beta }$
are functions of the space coordinates only; the inverse metric tensor is
\begin{equation*}
\gamma ^{\alpha \beta }=a^{\alpha \beta }-b^{\alpha \beta }\;t-c^{\alpha
\beta }\;t^{n}+\cdots ,
\end{equation*}%
so that $\gamma _{\alpha \beta }\gamma ^{\beta \gamma }=\delta _{\alpha
}^{\gamma }$. We note that $a_{\alpha \beta }a^{\beta \gamma }=\delta
_{\alpha }^{\gamma }$ and the indices of $b_{\alpha \beta },c_{\alpha \beta
} $ are raised by $a^{\alpha \beta }$. The extrinsic curvature, its
derivatives, and its contractions are, to leading order,
\begin{eqnarray*}
K_{\alpha \beta } &=&\frac{\partial \gamma _{\alpha \beta }}{\partial t}%
=b_{\alpha \beta }+c_{\alpha \beta }n\;t^{n-1}+\cdots , \\
K_{\beta }^{\alpha } &=&b_{\beta }^{\alpha }+nc_{\beta }^{\alpha
}\;t^{n-1}+\cdots , \\
K &=&b+nc\;t^{n-1}+\cdots , \\
\frac{\partial K_{\beta }^{\alpha }}{\partial t} &=&n(n-1)c_{\beta }^{\alpha
}\;t^{n-2}+\cdots , \\
\frac{\partial K}{\partial t} &=&n(n-1)c\;t^{n-2}+\cdots , \\
K_{\alpha }^{\beta }K_{\beta }^{\alpha } &=&b_{\mu \nu }b^{\mu \nu
}+2nb_{\mu \nu }c^{\mu \nu }\;t^{n-1}+\cdots , \\
KK_{\beta }^{\alpha } &=&bb_{\beta }^{\alpha }+n(cb_{\beta }^{\alpha
}+bc_{\beta }^{\alpha })\;t^{n-1}+\cdots
\end{eqnarray*}

The components of the Ricci tensor are
\begin{eqnarray*}
R_{0}^{0} &=&-\frac{1}{2}\frac{\partial K}{\partial t}-\frac{1}{4}K_{\alpha
}^{\beta }K_{\beta }^{\alpha }=-\frac{n(n-1)c}{2}\;t^{n-2}-\frac{1}{4}b^{\mu
\nu }b_{\mu \nu }-\frac{n}{2}c^{\mu \nu }b_{\mu \nu }\;t^{n-1}+\cdots , \\
R_{\alpha }^{0} &=&\frac{1}{2}(K_{\alpha \;;\beta }^{\beta }-K_{;\alpha })=%
\frac{1}{2}(b_{\alpha ;\beta }^{\beta }-b_{;\alpha })+\frac{n}{2}(c_{\alpha
;\beta }^{\beta }-c_{;\alpha })\;t^{n-1}+\cdots , \\
R_{\beta }^{\alpha } &=&-Q_{\beta }^{\alpha }-\frac{1}{4}KK_{\beta }^{\alpha
}-\frac{1}{2}\frac{\partial K_{\beta }^{\alpha }}{\partial t}=-\frac{n(n-1)}{%
2}c_{\beta }^{\alpha }\;t^{n-2}-[\frac{1}{4}bb_{\beta }^{\alpha }+P_{\beta
}^{\alpha }]-\frac{n}{4}[cb_{\beta }^{\alpha }+bc_{\beta }^{\alpha
}]\;t^{n-1}+\cdots ,
\end{eqnarray*}%
where $Q_{\alpha \beta }$ is the Ricci tensor associated with $\gamma
_{\alpha \beta },$ and $P_{\alpha \beta }$ is the Ricci tensor associated
with $a_{\alpha \beta }$. In the synchronous reference frame
\begin{equation*}
g_{00}=1,\qquad g_{0\alpha }=0.
\end{equation*}%
The 4-velocity $u^{i}=(u^{0},u^{\alpha })$ has $u^{0}=u_{0}$ and
\begin{equation*}
1=u_{i}u^{i}=u_{0}^{2}-(a_{\alpha \beta }+b_{\alpha \beta }\;t+c_{\alpha
\beta }\;t^{n}+\cdots )u^{\alpha }u^{\beta }.
\end{equation*}%
The stress energy tensor for a perfect fluid
\begin{equation*}
T_{j}^{i}=8\pi \lbrack (\rho +p)u^{i}u_{j}-p\delta _{j}^{i}]
\end{equation*}%
gives (to lowest order)
\begin{eqnarray*}
T_{0}^{0} &=&(\rho +p)u_{0}^{2}-p\approx \rho , \\
T_{\alpha }^{0} &=&(\rho +p)u_{0}u_{\alpha }\approx (\rho +p)u_{\alpha }, \\
T_{\beta }^{\alpha } &=&(\rho +p)u^{\alpha }u_{\beta }-p\delta _{\beta
}^{\alpha }\approx -p\delta _{\beta }^{\alpha }.
\end{eqnarray*}

From the Einstein equations
\begin{equation*}
R_{j}^{i}-\frac{1}{2}R\delta _{j}^{i}=8\pi T_{j}^{i},
\end{equation*}%
we use the $\binom{0}{0}$ component to calculate the energy density and the
trace of the $\binom{\alpha }{\beta }$ equations to calculate the pressure.
The 4-velocity can be calculated then from the $\binom{0}{\alpha }$
component. Also we use the $\binom{\alpha }{\beta }$ components to get
restrictions on the arbitrary functions of $a_{\alpha \beta },b_{\alpha
\beta },c_{\alpha \beta }$. We have
\begin{equation*}
16\pi \rho =\frac{1}{2}(R_{0}^{0}-R_{\alpha }^{\alpha })=\left( P+\frac{%
b^{2}-b^{\mu \nu }b_{\mu \nu }}{4}\right) -\frac{n}{2}(b^{\mu \nu }c_{\mu
\nu }-bc)\;t^{n-1}+\cdots ,
\end{equation*}%
while from the trace of the $\binom{\alpha }{\beta }$ component we get
\begin{equation*}
16\pi p=\frac{1}{2}(R_{0}^{0}+\frac{1}{3}R_{\alpha }^{\alpha })=-\frac{%
2n(n-1)c}{3}\;t^{n-2}-\frac{3b^{\mu \nu }b_{\mu \nu }+b^{2}+4P}{12}-\frac{n}{%
2}(b^{\mu \nu }c_{\mu \nu }+\frac{bc}{3})\;t^{n-1}+\cdots
\end{equation*}

The Ricci scalar is
\begin{equation*}
R=R_{i}^{i}=-n(n-1)ct^{n-2}-\frac{b_{\mu \nu }b^{\mu \nu }+b^{2}+4P}{4}-%
\frac{n}{2}(b^{\mu \nu }c_{\mu \nu }+bc)\;t^{n-1}+\cdots ,
\end{equation*}%
the $\binom{\alpha }{\beta }$ components of the Einstein equations give
\begin{equation*}
R_{\beta }^{\alpha }=\frac{1}{2}(R-16\pi p)\delta _{\beta }^{\alpha },
\end{equation*}%
and therefore
\begin{eqnarray*}
&&-\frac{n(n-1)}{2}c_{\beta }^{\alpha }\;t^{n-2}-\left( \frac{1}{4}bb_{\beta
}^{\alpha }+P_{\beta }^{\alpha }\right) -\frac{n}{4}(cb_{\beta }^{\alpha
}+bc_{\beta }^{\alpha })\;t^{n-1}+\cdots \\
&=&-\frac{n(n-1)c}{6}\delta _{\beta }^{\alpha }\;t^{n-2}-\delta _{\beta
}^{\alpha }\left( \frac{b^{2}+4P}{12}\right) -\delta _{\beta }^{\alpha }%
\frac{nbc}{6}\;t^{n-1}+\cdots
\end{eqnarray*}

From the $O(t^{n-2})$ terms, we get the constraints
\begin{equation}
c_{\beta }^{\alpha }=\frac{c}{3}\delta _{\beta }^{\alpha }\qquad or\qquad
c_{\alpha \beta }=\frac{c}{3}a_{\alpha \beta },  \label{c}
\end{equation}%
and so the components $c_{\alpha \beta }$ are all but one determined from $%
a_{\alpha \beta }$.

Since $\gamma _{\alpha \beta }$ is to order $t^{n}$, the extrinsic curvature
will be $O(t^{n-1})$ and the Ricci components $R_{0}^{0},\ R_{\beta
}^{\alpha }$ will be $O(t^{n-2})$. Hence this will be the only restriction
on $a_{\alpha \beta },b_{\alpha \beta },c_{\alpha \beta }$. Therefore, we
have $6+6+1=13$ independent functions. Subtracting the $4$ coordinate
covariances which may still be used to remove four functions, leaves $9$
independent arbitrary functions of the three space coordinates on a surface
of constant $t$ time.

\bigskip\ This is the maximal number of independent arbitrary spatial
functions expected in a local representation of part of the general solution
of Einstein's equations near a sudden singularity. In general, we expect
there will be $6\times g_{_{\alpha \beta }}$ and $6\times \dot{g}_{_{\alpha
\beta }}$, plus $3$ free velocity components, $u_{\alpha },$and $p$ and $%
\rho $, giving a total of $17$ independent functions. We can remove four of
these by using the $R_{a}^{0}$ constraints and four more by using the
general coordinate covariances. This leaves a total of $9$ free functions
expected in the general solution for the metric locally. If an equation of
state had been assumed to relate the pressure to the density this number
would have been reduced by $1$ to $8$. \newline
\ \ \ \ \ \ \ We have found that the pressure is $O(t^{n-2})$, so $\rho
+p=O(t^{n-2}),$ and from
\begin{equation*}
R_{\alpha }^{0}=8\pi (\rho +p)u_{0}u_{\alpha }\approx 8\pi (\rho
+p)u_{\alpha },
\end{equation*}%
since $R_{\alpha }^{0}\approx O(t^{0}),$ we get that $u_{\alpha }\approx
O(t^{-(n-2)}),$ and the full expansion is
\begin{equation*}
u_{\alpha }=-\frac{3(b_{\alpha ;\beta }^{\beta }-b_{;\alpha })}{2n(n-1)c}%
\;t^{2-n}\equiv \Phi _{\alpha }\;t^{2-n},\qquad 0<2-n<1.
\end{equation*}

The equation for the stress energy conservation $T_{\ \ ;i}^{ij}=0$ is split
into time and spatial components as follows:
\begin{eqnarray*}
T_{\ \ ;i}^{i0} &=&T_{\ \ ;\alpha }^{\alpha 0}+\partial _{t}T^{00}+\frac{1}{2%
}(KT^{00}+K_{\alpha \beta }T^{\alpha \beta })=0, \\
T_{\ \ ;i}^{i\beta } &=&T_{\ \ ;\alpha }^{\alpha \beta }+\partial
_{t}T^{0\beta }+\frac{1}{2}KT^{0\beta }+K_{\alpha }^{\beta }T^{0\alpha }=0,
\end{eqnarray*}%
where
\begin{equation*}
T^{00}=\rho ,\qquad T^{0\alpha }=(\rho +p)u^{\alpha },\qquad T^{\alpha \beta
}=p\gamma ^{\alpha \beta }.
\end{equation*}%
The time component gives
\begin{equation}
\lbrack (\rho +p)u^{\alpha }]_{;\alpha }+\partial _{t}\rho +\frac{1}{2}%
K(\rho +p)=0.  \label{p}
\end{equation}%
To lowest order, the first term of eq. (\ref{p}) is constant ($O(t^{0})$).
The other two terms to lowest order are $O(t^{n-2})$; therefore, to lowest
order the equation is $O(t^{n-2})$. Hence the $O(t^{n-2})$ term of $\partial
_{t}\rho $ must cancel the $O(t^{n-2})$ term of $\frac{1}{2}K(\rho +p)$. The
first term is
\begin{equation*}
16\pi (\rho +p)u^{\alpha }=-a^{\alpha \beta }(b_{\beta ;\mu }^{\mu
}-b_{;\beta })-na^{\alpha \beta }(c_{\beta ;\mu }^{\mu }-c_{;\beta
})t^{n-1}+\cdots
\end{equation*}%
The second is
\begin{equation*}
16\pi \partial _{t}\rho =-\frac{n(n-1)}{2}(b_{\mu \nu }c^{\mu \nu
}-bc)t^{n-2}+\cdots ,
\end{equation*}%
and the third is
\begin{equation*}
\frac{1}{2}K(16\pi \rho +16\pi p)=-\frac{n(n-1)bc}{3}t^{n-2}+\cdots .
\end{equation*}%
Therefore this equation will be satisfied to order $O(t^{n-2})$ if
\begin{equation*}
\frac{n(n-1)}{2}(b_{\mu \nu }c^{\mu \nu }-bc)+\frac{n(n-1)bc}{3}=0,
\end{equation*}%
that is, by
\begin{equation*}
b_{\mu \nu }c^{\mu \nu }=\frac{1}{3}bc,
\end{equation*}%
which is indeed satisfied because of the constraint eq. (\ref{c}) on $%
c_{\alpha \beta }$ found earlier. Therefore, the time component of $T_{\ \
;i}^{ij}=0$ does not give any new constraints on the terms of the series.

Similarly, the spatial part of the stress-energy conservation is expressed
as
\begin{equation*}
\gamma ^{\alpha \beta }p_{;\alpha }+\partial _{t}[(\rho +p)u^{\beta }]+\frac{%
1}{2}K(\rho +p)u^{\beta }+K_{\alpha }^{\beta }(\rho +p)u^{\alpha }=0.
\end{equation*}%
The lowest order of the first two terms is $O(t^{n-2})$ while the lowest
order of the last two terms is $O(t^{0})$ and so they are neglected. For the
first term,
\begin{equation*}
16\pi \gamma ^{\alpha \beta }p_{;\alpha }=-\frac{2n(n-1)}{3}a^{\alpha \beta
}c_{;\alpha }t^{n-2}+\cdots ,
\end{equation*}%
while for the second,
\begin{equation*}
16\pi \partial _{t}[(\rho +p)u^{\beta }]=-n(n-1)a^{\beta \nu }(c_{\nu ;\mu
}^{\mu }-c_{;\nu })t^{n-2}+\cdots .
\end{equation*}%
Hence, we must have
\begin{equation*}
\frac{2n(n-1)}{3}a^{\alpha \beta }c_{;\alpha }+n(n-1)a^{\beta \nu }(c_{\nu
;\mu }^{\mu }-c_{;\nu })=0
\end{equation*}%
or
\begin{equation*}
c_{\alpha ;\beta }^{\beta }=\frac{1}{3}c_{;\alpha }
\end{equation*}%
which again is satisfied by the restriction of eq. (\ref{c}) on $c_{\alpha
\beta }$. Hence, both the temporal and the spatial components of the
conservation of the stress energy tensor are satisfied to lowest order by
the constraint on $c_{\beta }^{\alpha }$ that we have found already and
there remain $9$ independent functions in our series solution.

As an aside we note that he speed of sound is
\begin{equation*}
v_{s}^{2}=\frac{dp}{d\rho }=\frac{\dot{p}}{\dot{\rho}}=\frac{4(2-n)c}{%
3(bc-b_{\alpha \beta }c^{\alpha \beta })}t^{-1}
\end{equation*}%
which goes to infinity as $t\rightarrow 0$ which is unsurprising given the
divergence of the pressure at the sudden singularity$.$

\subsection{ Puiseux and logarithmic series}

\bigskip We shall now discuss some other series expansions around the
singularity. Consider first a Puiseux series of the form
\begin{equation*}
a(t)=\sum_{j=0}^{\infty
}a_{j}t^{j}+t^{n}[b_{0}+t^{1/q}b_{1}+t^{2/q}b_{2}+\cdots ]
\end{equation*}%
as an asymptotic form for the Friedmann-Lema\^{\i}tre equations instead of
eq. (\ref{sol2}). This suggests an ansatz for the quasi-isotropic form of
the general inhomogeneous metric of the form
\begin{equation*}
\gamma _{\alpha \beta }=a_{\alpha \beta }+b_{\alpha \beta }\;t+c_{\alpha
\beta }\;t^{n}+\psi _{\alpha \beta }\;t^{m}+\omega _{\alpha \beta
}\;t^{r}+\cdots ,
\end{equation*}%
with
\begin{equation*}
1<n<m<r<2.
\end{equation*}%
If we perform the same analysis we will find again that $a_{\alpha \beta }$
and$\ b_{\alpha \beta }$ are arbitrary and
\begin{equation*}
c_{\alpha \beta }=\frac{c}{3}a_{\alpha \beta },\qquad \mbox{and}\qquad \psi
_{\alpha \beta }=\frac{\psi }{3}a_{\alpha \beta },\qquad \mbox{and}\qquad
\omega _{\alpha \beta }=\frac{\omega }{3}a_{\alpha \beta },
\end{equation*}%
with $c,\ \psi ,\ \omega $ again arbitrary. The pressure and the density
will be
\begin{eqnarray}
16\pi \rho &=&16\pi \rho _{0}+\frac{nbc}{3}\;t^{n-1}+\frac{mb\psi }{3}%
\;t^{m-1}+\frac{rb\omega }{3}\;t^{r-1}+\cdots ,  \label{rho1} \\
16\pi p &=&\ -\frac{2n(n-1)c}{3}\;t^{n-2}\left[ 1+\frac{m(m-1)\psi }{n(n-1)c}%
\;t^{m-n}+\frac{m(m-1)\omega }{n(n-1)c}\;t^{r-n}+\cdots \right] ,
\label{pee1}
\end{eqnarray}%
and
\begin{equation*}
m-n>0\qquad r-n>0.
\end{equation*}%
Therefore we have the same phenomenon as with the series that was described
in the previous section. If we keep adding terms of the form $t^{s}$ with $%
1<s<2$ we will getting one more arbitrary function so if we add an infinite
number of terms we will have an infinite number of functions. The reason
that they are irrelevant is that they do not alter the nature of the
singularity in its neighborhood as can be seen from the density and pressure
eqs. (\ref{rho1})-(\ref{pee1}). In other words, all three forms
\begin{eqnarray*}
\gamma _{\alpha \beta } &=&a_{\alpha \beta }+b_{\alpha \beta }\;t+c_{\alpha
\beta }\;t^{n}+\cdots , \\
\gamma _{\alpha \beta } &=&a_{\alpha \beta }+b_{\alpha \beta }\;t+c_{\alpha
\beta }\;t^{n}+\psi _{\alpha \beta }\;t^{m}+\omega _{\alpha \beta
}\;t^{r}+\cdots , \\
\gamma _{\alpha \beta } &=&a_{\alpha \beta }+b_{\alpha \beta }\;t+c_{\alpha
\beta }\;t^{n}+\psi _{\alpha \beta }\;t^{m}\ln (t)\cdots
\end{eqnarray*}%
give
\begin{equation*}
\lim_{t\rightarrow 0}\rho =\rho _{0},\qquad \mbox{and}\qquad 16\pi
\lim_{t\rightarrow 0}p=-\lim_{t\rightarrow 0}\frac{2n(n-1)c}{3}\;t^{n-2}
\end{equation*}%
and hence the same behaviour on approach to the sudden singularity as $%
t\rightarrow 0$. Thus the number of relevant independent functions is $%
9=6+6+1-4$ which is the number required for a specification of the general
solution.

\section{Discussion}

If we had sought a power-series expansion of the metric with time dependence
dictated by the form of the radiation-dominated Friedmann-Lema\^{\i}tre
universe on approach to an initial big bang curvature singularity as $%
t\rightarrow 0$ then it would have taken the form of eq. (\ref{met}) with

\begin{equation}
\gamma _{\alpha \beta }=\ ta_{\alpha \beta }+t^{2}b_{\alpha \beta }+\ ...
\label{rad}
\end{equation}%
with $a_{\alpha \beta }$, and $b_{\alpha \beta }$ spatial functions \cite%
{LL,KL,KL2} However, at most 3 arbitrary spatial functions in the series
expansion are left free in this solution. A general solution of this form,
with a radiation equation of state requires 8 independently arbitrary
functions of space. The reason for the non-generic form of (\ref{rad}) for
the radiation fluid is its quasi-isotropic character. On approach to a
general relativistic cosmological singularity the anisotropic shear and
3-curvatures dominate the dynamics and destroy the quasi isotropic character
of a radiation-dominated universe.

By contrast, our results are quite different. We find that on approach to a
late-time sudden singularity, where the density, expansion rate, and metric
remain finite, it is possible for a quasi-isotropic solution to be part of
the general solution of the Einstein equations. This solution has no
equation of state and is characterised by nine independently arbitrary
spatial functions. On approach to a sudden singularity the mean scale factor
and the shear approach constant values, the chaotically anisotropic degrees
of freedom which dominate at an initial curvature singularity are frozen
out, and anisotropic velocities grind to a halt because of the divergent
pressure and inertia. The situation is far simpler than for a initial vacuum
or $p<\rho $ dominated fluid singularity. Comparable simplicity is achieved
in the non-singular late-time approach to a quasi-isotropic de Sitter
universe in the presence of a cosmological constant or a $p=-\rho $ fluid,
shown by Starobinsky \cite{no hair}. We have shown that there is an
analogous 'no hair' behaviour on approach to a late-time sudden singularity
and the general structure of such a finite-time singularity admits a simple
quasi-isotropic form.

\end{document}